\documentclass[twocolumn]{aastex62}
\usepackage{lineno}

\usepackage{natbib}
\usepackage{float}
\usepackage{amsmath}
\usepackage{gensymb}
\graphicspath{{./}{figures/}}

\accepted{July 2025}
\submitjournal{ApJ}

\shorttitle{G18.9-1.1 NS Velocity}
\shortauthors{Holland-Ashford et. al.}

\begin{document}

\title{Proper Motion of the Neutron Star in the Supernova Remnant G18.9$-$1.1}
\correspondingauthor{Tyler Holland-Ashford}
\email{tyler.e.holland-ashford@nasa.gov}
\author{Tyler Holland-Ashford}
\affil{Astrophysics Science Division, NASA Goddard Space Flight Center Greenbelt, MD 20771, USA}
\author{Brian Williams}
\affil{Astrophysics Science Division, NASA Goddard Space Flight Center Greenbelt, MD 20771, USA}
\author{Patrick Slane}
\affil{Center for Astrophysics $|$ Harvard \& Smithsonian, 60 Garden St, Cambridge MA 02138, USA}
\author{Xi Long}
\affil{Department of Physics, The University of Hong Kong, Pokfulam Road, Hong Kong}


\begin{abstract}
In this paper, we present the first direct measurement of the proper motion of the neutron star (NS) in the supernova remnant (SNR) G18.9$-$1.1 using a 15-year {\it Chandra} baseline. After correcting the observations' astrometric solutions using reference Gaia stars' positions, we measure a total proper motion of 24.7 $\pm$ 6.8 mas yr$^{-1}$ at an angle of $336^\circ \pm$ 16$^\circ$ east of north. Using the distance estimates from literature of 2.1~kpc and 3.8~kpc, this proper motion corresponds to Galactic rotation-corrected transverse velocities of 264d$_{2.1}$ $\pm$ 79 km s$^{-1}$ and 474d$_{3.8}$ $\pm$ 129 km s$^{-1}$, respectively. Our power ratio method analysis of SNR ejecta slightly favors the higher velocity, as multipole moments calculated from the back-evolved center using the farther distance are more consistent with values from other CCSNRs. The NS's motion is directly opposite the motion of bulk ejecta in G18.9$-$1.1, providing yet more evidence that NS kicks are generated via a conservation of momentum-like process between the NS and the ejecta, as has been observed in other SNRs.

\end{abstract}

\keywords{compact objects: neutron stars -- supernovae: individual (G018.9--1.1) -- X-ray astronomy: X-ray point sources -- ISM: supernova remnants}

\section{Introduction}
Neutron stars (NS) are compact objects formed in some core-collapse supernovae (CCSNe) and have been observed to have typical spatial velocities of a few hundred km s$^{-1}$ \citep{hobbs05,verbunt17,igoshev20}. As these fast NS ``kicks'' are greater than any velocity that would arise from the disruption of a binary (v$_{\rm NS}\approx100$ km s$^{-1}$; \citealt{lai01}), it is thought that they are generated as a result of instabilities in the supernova explosion. Indeed, there is growing evidence that most NS are accelerated in a direction opposite the bulk ejecta motion in a conservation of momentum-like process \citep{scheck06,wongwathanarat13,janka17,me17,katsuda18,burrows23}. However, there remains merit to the theory that NSs can be kicked in a direction opposite the bulk neutrino emission \cite{fryer06}, particularly for CCSNe with the lowest progenitor masses, although such kicks are thought to be $\lesssim$100 km s$^{-1}$\citep{coleman22,burrows23}.

To investigate the origins of these kicks and further our understanding of SN explosion processes, we can study NSs that are directly associated with supernova remnants (SNRs). The age of and distance to such NSs matches that of the SNR, and the properties of such NSs can be compared to those of the SNR and its progenitor: e.g., energy, explosion type, ejecta asymmetries, shock evolution, progenitor mass, and progenitor binarity (e.g., in \citealt{reynolds17,me20a,zhou20,narita23,kim24}. Specifically, young SNRs with strong thermal emission from shock-heated ejecta are prime targets for such studies, as the X-ray emission will more directly probe SN mechanisms rather than properties of the surrounding medium swept-up by the forward shock (e.g., see reviews by \citealt{weiss06b,vink12}). Furthermore, young, SNR-embedded NSs are most easily and consistently detected in X-ray observations. Certain types of NSs emit at only X-ray wavelengths (e.g., central compact objects; \citealt{deluca17}), and various environmental effects (e.g., foreground absorption, SNR brightness; \citealt{kaspi00,sett21}) can hinder the detection of NSs still embedded within their SNR at optical and radio wavelengths. 

Robust NS velocities have typically been obtained using one of two methods. In the first method, the SN explosion site---i.e., the NS's birth site---is found via identifying numerous individual SNR ejecta knots, measuring the motion of each, and back-evolving them to a central location \citep{fesen06,winkler09,banovetz21}. The NS velocity can then be estimated as the distance between the SNR explosion site and the NS's current location divided by the NS's age. Due to the difficulty of obtaining precise proper motions for many knots, this technique has so far only been done for a handful of NS-SNR systems. For example, although \cite{tsuchioka21} back-evolved five ejecta knots to a central location to conclude that the SNR G350.1$-$0.3 is about 655~yr old, the exact center of the SNR has large error bars---in part due to the unknown deceleration of each ejecta knot.

The second method is to measure the proper motion of the NS directly. This technique is difficult due to NSs' relatively small proper motions ($\sim$0.\arcsec03 yr$^{-1}$ for a source 3~kpc away moving at a typical velocity 400 km s$^{-1}$; \citealt{hobbs05}). At X-ray wavelengths, only {\it Chandra} has the spatial resolution necessary for these measurements, and even then a baseline of $\gtrsim$10~years is necessary. However, in order to obtain NS positions that are accurate to better than {\it Chandra}'s $\sim$0.\arcsec5 astrometric accuracy\footnote{\url{https://cxc.harvard.edu/cal/ASPECT/celmon/}}, the astrometric solution of each observation must be corrected via using multiple registration sources. So far, the proper motion of $\sim$8 NSs associated with SNRs exhibiting significant thermal emission have been made \citep{temim17,mayer20,halpern15,mayer21,long22,me24}, with another $\sim$7 measurements made for NSs associated with SNRS not exhibiting thermal emission \cite{ho20,devries21,vanetten12,halpern15,dzib21,shternin19}. These studies were able to obtain total motion uncertainties of 0.\arcsec05--0.\arcsec2 corresponding to velocity uncertainties of $\lesssim$250 km s$^{-1}$ over their 10--20-year baselines. Although most NS proper motion measurements are made using {\it Chandra}, there have been a few measurements using radio data (e.g., \citealt{zeiger08,shternin19,dzib21}. However, none of these sources have associated SNRs that still are dominated by thermal emission from ejecta.

Finally, there are a few other methods of measuring NS proper motions, although these are less reliable. In the ``geometric method'', the distance from the NS's current location to the center of a SNR's X-ray or radio profile is measured and then divided by the age of the SNR to obtain a velocity estimate (e.g., \citealt{tullmann10}). Velocities obtained this way have large, unknown uncertainties as it is not guaranteed that the geometrical center of the SNR matches with its true explosion site as SNRs can be highly asymmetric (e.g., \citealt{lopez09b, me17}). Alternatively, the pulsar velocity can be estimated by relating it to the SNR shock velocity or by solving for pressure equilibrium conditions between any pulsar wind and the surrounding medium (e.g., \citealt{frail96}). However, similar to the geometric method, both of these methods require additional assumptions that introduce unknown errors to the final velocity estimate.

The SNR G18.9$-$1.1 (AKA: G18.95$-$1.1; Hereafter: G18.9) is a $\sim$5~kyr-old SNR whose spectra can be well-fit by an absorbed thermal plasma with N$_H$ = (3--10) $\times 10^{21}$ cm$^{-2}$ and super-solar ejecta abundances \citep{fuerst97,harrus04, bykov22}. The X-ray emission from G18.9 is clearly asymmetric, with the bulk of the ejecta emission present in the SE. It has a few different reported distance estimates: 2.1 $\pm$ 0.4~kpc \citep{ranasinghe19}, 1.8 $\pm$ 0.2~kpc \citep{shan18}, 3.8 $\pm$ 0.4~kpc \citep{zhou23}, and $>$2~kpc \citep{lee20}. Throughout this paper, we consider both d$_{2.1}$=2.1~kpc and d$_{3.8}$=3.8~kpc. The shorter distance favors a SNR age of $\sim$5.5~kyr and lower ionization timescales, while the larger distance favors a $\sim$10~kyr age and a plasma closer to collisional ionization equilibrium (CIE); previous analysis of {\it ROSAT} and {\it eROSITA} data has produced fits consistent with both scenarios \citep{harrus04,bykov22}. The identified NS (CXOU J182913.1--125113) has a reported geometric velocity of $\sim$700-960d$_{2}$ km s$^{-1}$ \citep{tullmann10}.

In this paper, we measure the proper motion of the NS in G18.9 using a baseline of 15~years between {\it Chandra} X-ray observations. We correct the astrometry of the {\it Chandra} observations using multiple reference Gaia sources in order to obtain a robust measurement of the NS's position at each epoch. Finally, we compare the derived NS velocity to SNR ejecta asymmetries to help place constraints on explosion processes and mechanisms responsible for generating the NS kick.

Our paper is formatted as follows: in Section~\ref{sec:methods}, we present the {\it Chandra} observations used and the process of using reference Gaia sources to correct their astrometry. In Section~\ref{sec:results}, we present our NS proper motion measurement and its relation to bulk ejecta motion in G18.9. In Section~\ref{sec:conc}, we present our conclusions and discuss future related work.

\begin{figure*}
\begin{center}
\includegraphics[width=0.48\textwidth]{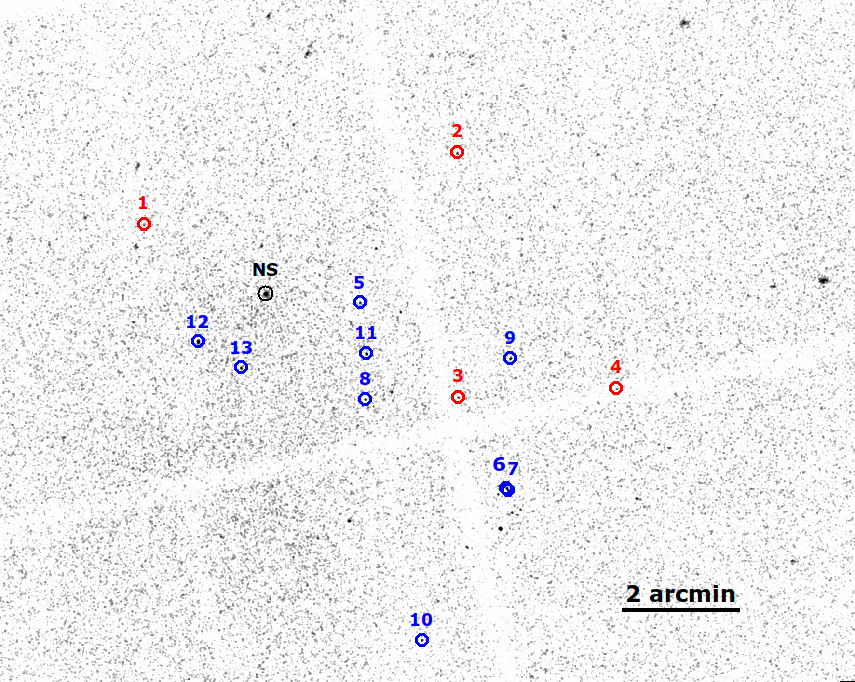}
\includegraphics[width=0.48\textwidth]{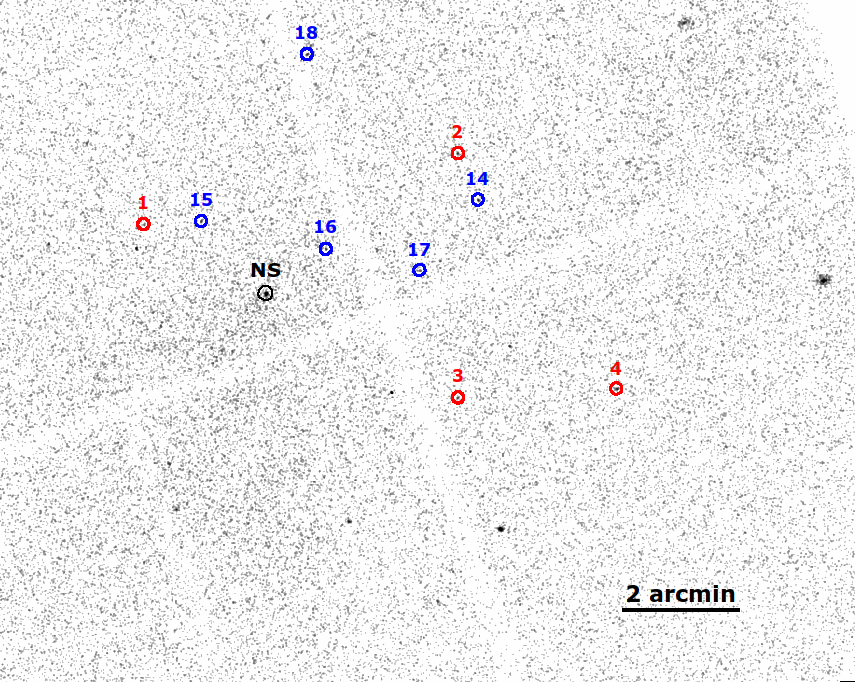}
\end{center}
\vspace{-5mm}
\caption{\footnotesize{{\it Chandra} images of the NS and surrounding region: ObsID~10098 on the left, and observation created from merging ObsID 26656 and 29478 on the right. The NS is labeled in black and the sources used for astrometric calibration are in either red (for sources detected in both epoch observations) or blue. All circles are arbitrarily sized and don't reflect the point source centroid uncertainties. }}
\label{fig:source_images}
\end{figure*}

\section{Observations and Data Analysis}
\label{sec:methods}
\subsection{{\it Chandra} Observations}
The NS in G18.9 was observed with {\it Chandra} in September, 2009 (ObsID: 10098; PI: Tuellmann) for 44.5~ks. Additional {\it Chandra} time was awarded in Cycle 24, and observations were taken in July, 2024 in two separate observations: ObsID 26656 for 20.8~ks and ObsID 29478 for 23.2~ks. This created a 15-year baseline with which to measure the NS’s motion. 

\subsection{Merging the 2024 Observations}
\label{subsec:merging2024}
First, we reprojected and merged the two 2024 observations to make full use of the combined 44~ks observation time. We used the {\sc CIAO} tool \texttt{wavdetect} with keyword \texttt{sigthresh}=1e-4 to find faint sources in each observation, and then manually selected 13 sources that were detected in both observations, isolated from other sources, and circular (i.e., not too smeared by off-axis PSF). We then solved for a transformation matrix to match the astrometric solution of one observation to that of the other: 
\begin{equation}
    \begin{pmatrix}
    x'\\
    y'
    \end{pmatrix}
    = 
    \begin{pmatrix}
    {\rm r} \cos\theta & -{\rm r}\sin\theta \\
    {\rm r} \sin\theta  & {\rm r}\cos\theta 
    \end{pmatrix}
     \begin{pmatrix}
    x\\
    y
    \end{pmatrix}
    +
     \begin{pmatrix}
    \Delta x\\
    \Delta y
    \end{pmatrix},
\end{equation}
where (x,y) are input pixel coordinates of point sources,  (x',y') are the transformed coordinates, r represents a linear stretching of the image, $\theta$ is rotation, and ($\Delta x$, $\Delta y$) represents horizontal and vertical translational shifts in units of pixels. The rotation is performed around the aimpoint of the observation, and all parameters are allowed to vary freely in our fit. 

To solve for the transformation matrix, we used a least squares algorithm (Python function \texttt{scipy.optimize.\\least\_squares}), weighting each point source by its centroid uncertainty. Specifically, we used the `\texttt{soft\_l1}' loss function option \citep{triggs00} that weighs outliers with a linear rather than a quadratic penalty and produces a more accurate overall transformation \citep{long22}. Finally, we used {\sc CIAO}'s \texttt{wcs\_update} tool to reproject the aspect solution of the shorter observation (ObsID 26656: 20.8~ks) such that it matched that of the longer observation (ObsID 29478: 23.3~ks), and then we merged the two observations using \texttt{merge\_obs}. The best-fit parameters of this transformation matrix were [r, $\theta$, $\Delta x$, $\Delta y$] = [1.001, -0.038\degree, -0.107, -2.06].

\subsection{Astrometric Correction: 2009 and 2024}

For correcting the astrometry of ObsID 10098 and of the merged 2024 observation, we performed a similar process with a few key differences. 

1.) Due to the 15-year baseline between the observations, we needed to account for registration point sources' proper motions of up to $\sim$7 mas/yr, corresponding to total motions of up to $\sim$0.\arcsec1 over the 15-year baseline. When selecting sources to use for astrometric correction, we restricted our sample to those with a co-spatial source identified in the Gaia DR3 catalog\footnote{\url{https://www.cosmos.esa.int/web/gaia/dr3}} \citep{gaia16b,gaia23j,babusiaux23} that had reported proper motions. Figure~\ref{fig:source_images} shows the registration sources we used in each observation, and Table~\ref{table:pointsources} lists the corresponding Gaia designations, positions, and measured sky motions.

2.) Instead of correcting the aspect solution of one observation to that of the other, we instead corrected both images to an absolute reference frame: that given by Gaia DR3 point source positions. We chose this method because we identified 13 reference sources in ObsID 10098 and 9 registration sources in the merged 2024 observation, but only 4 of these were present in both epochs. Attempting to directly match the observations would result in a loss of over half of our registration sources.

3.) We accounted for the effects of PSF in obtaining point source positions in {\it Chandra} observations, following the methods of past papers (e.g., \citealt{mayer20,long22,me24}---see Section 2.4 of the latter paper for the detailed step-by-step process). We used the {\sc CIAO} Chandra Ray Tracer (ChaRT) tool\footnote{\url{https://cxc.harvard.edu/ciao/PSFs/chart2/index.html}} to  generate event files and PSFs for each point source, binning these event files by 1/8 to better enable sub-pixel localization. Finally, we followed the steps outlined in the ``Accounting for PSF Effects in 2D Image Fitting'' \texttt{CIAO} Sherpa \citep{freeman01} thread\footnote{\url{https://cxc.cfa.harvard.edu/sherpa/threads/2dpsf/}} to fit the centroid of each point source. These positions and their associated uncertainties are more accurate than positions reported via \texttt{wavdetect}.

Finally, for each {\it Chandra} observation, we solved for a transformation matrix to correct its astrometry to that of the true Gaia reference frame as in Section~\ref{subsec:merging2024}. After obtaining the transformation matrix for each observation, we applied it to the detected NS position in each observation to obtain the astrometry-corrected NS positions. The best-fit parameters of these transformation matrix were [r, $\theta$, $\Delta x$, $\Delta y$] = [0.998, 0.00542\degree, -0.352, -0.116] for the 2009 observation and [1.000, -0.0231\degree, -0.144, -3.400] for the merged 2024 observation.

\subsection{Uncertainties on Final NS Positions}
To obtain a final uncertainty term for the NS positions in each epoch, we had to account for the following sources of uncertainty: 1.) the best-fit centroid uncertainty of the NS, 2.) the best-fit centroid uncertainties of all the registration sources, and 3.) the uncertainty associated with the transformation matrix. Error term \#1, the centroid uncertainty of the NS, is output by the 2D-PSF fitting. Error term \#2, the total registration source uncertainty, is found via inverse variance weighting of the individual registration source centroid uncertainties:
\begin{equation}
     \sigma_{\rm PSs, tot} = \left(\frac{1}{\sum_i^N (1/\sigma_{i}^2)}\right)^{0.5}.
\end{equation}
Error term \#3, the uncertainty in the transformation matrix (i.e., the accuracy of the astrometric correction) is equal to the weighted average of the difference (d) between the ``true'' and corrected registration source locations:
\begin{equation}
    \sigma_{\rm Res, tot} = \frac{\sum_i^N (d/\sigma_{i}^2)}{\sum_i^N (1/\sigma_{i}^2)}.
\end{equation}
Both of the previous error terms are calculated using inverse weighting because we weighed each registration source by its uncertainty when solving for the transformation matrix. Thus, the terms with smaller errors dominate, and all the error terms combine to obtain a smaller uncertainty than any individual registration source's centroid uncertainty.

The total uncertainty in NS position at each epoch is calculated by summing each of the above three error terms in quadrature, for RA and Dec separately. We ignored the uncertainties associated with the ``true'' Gaia source positions ($\lesssim$0.\arcsec001) as they were negligible compared to the detected point source uncertainties (0.06--0.\arcsec3) and astrometric correction accuracy ($\sim$0.\arcsec015). We also ignored any uncertainties due to the parallax difference between the registration sources and the NS, as they are $\lesssim$0.\arcsec0009 and thus negligible.

\begin{deluxetable*}{crrrrrcc}[!t]
\tablecolumns{7}
\tablewidth{0pt} 
\tablecaption{Registration Point Sources \label{table:pointsources}}
\tablehead{ \colhead{Source} & \colhead{Gaia DR3} & \colhead{RA}  &\colhead{Dec} & \colhead{$\mu_\alpha$cos$\delta$} & \colhead{$\mu_\delta$} & \colhead{In Epoch} &\colhead{In Epoch} \\ \colhead{} & \colhead{Designation} & \colhead{(deg)}  &\colhead{(deg)} & \colhead{(mas yr$^{-1}$)} & \colhead{(mas yr$^{-1}$)} & \colhead{2009?} &\colhead{2024?}  }
\startdata
1 & 4152918609988039552 & 277.339892 & -12.834033 & -1.36 & -1.88 & y & y \\
2 & 4152922488324001792 & 277.248815 & -12.813883 & -0.74 & 3.13 & y & y  \\
3 & 4152897818027107328 & 277.248695 & -12.883044 & -1.24 & -5.46 & y & y  \\
4 & 4152897680588956160 & 277.202523 & -12.880633 &  1.55 & -0.75 & y & y  \\
5 & 4152897994124153728 & 277.277209 & -12.856119 &  2.23 & -1.81 & y & n \\
6 & 4152897341308575104 & 277.234519 & -12.908753 & -0.44 & -2.16 & y & n \\
7 & 4152897337000674176 & 277.234204 & -12.909345 & -0.33 & -2.59 & y & n \\
8 & 4152894936126885376 & 277.275608 & -12.883592 & -0.03 & -5.41 & y & n \\
9 & 4152898028483938304 & 277.233389 & -12.871989 &  0.14 & -2.66 & y & n \\
10 & 4152894004096161408& 277.259090 & -12.951885 & -0.46 & -0.31 & y & n  \\
11 & 4152897891064389376& 277.275270 & -12.870741 &  6.40 & -5.19 & y & n \\
12 & 4152918399511237504& 277.324167 & -12.867184 &  1.89 & -1.55 & y & n \\
13 & 4152918365151486080& 277.311691 & -12.874713 & -1.11 & -3.12 & n & y \\
14 & 4152899024916423296& 277.242836 & -12.826857 &  1.01 & -1.35 & n & y \\
15 & 4152921526251381504& 277.323323 & -12.832997 &  1.98 & -1.736 & n & y \\
16 & 4152921461830603776& 277.287004 & -12.840980 &  0.38 & -2.31 & n & y \\
17 & 4152898200282632960& 277.259942 & -12.846921 & -0.73 & -1.80 & n & y \\
\enddata
\end{deluxetable*}

\subsection{Power Ratio Method}
To quantify the asymmetry of the ejecta in G18.9$-$1.1, we utilize the power ratio method (PRM; \citealt{lopez09a}), a multipole expansion technique used to quantitatively characterize the multipole moments of SNR ejecta emission (e.g., \citealt{lopez09b,me20a}). The multipole powers $P_{\rm m}$ of emission  are divided by the zeroeth order term $P_{0}$ to normalize PRM values and enable direct comparison between objects of different fluxes. Each multipole moment $P_{\rm m}$ reflects smaller and different measures of asymmetry; P$_1$ (the dipole moment) reflects bulk ejecta motion, P$_2$ (quadrupole moment) reflects ellipticity, and P$_3$ (octupole moment) reflects mirror asymmetry. We calculated the power ratios using a {\it ROSAT} image (RP500040) of the SNR, including 0.5--2.1~keV emission (dominated by emission from shocked ejecta) and centered on the SNR explosion site calculated by back-evolving the NS to its birth site. Both the PRM and the NS velocity measured here are 2D measurements, reflecting SNR asymmetry and NS velocity as projected into the plane of the sky.

\begin{figure*}
\begin{center}
\includegraphics[width=0.6\textwidth]{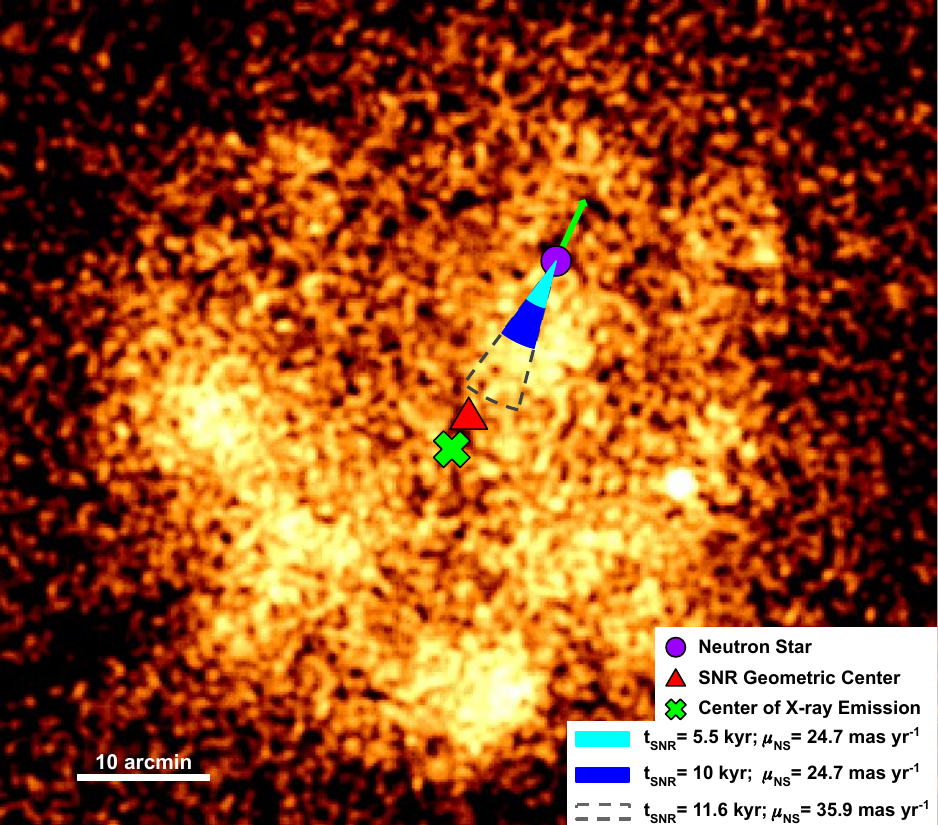}
\end{center}
\vspace{-5mm}
\caption{\footnotesize{A {\it ROSAT} 0.5--2.1~keV image (RP500040) of G18.9$-$1.1. The NS (purple) is moving with a proper motion of 24.7 mas yr$^{-1}$ angle of 336$^\circ$ east of north, as indicated by the green arrow. The cyan and blue cones show the NS's motion if back evolved for 5.5~kyr and 10.0~kyr (corresponding to distance estimates of 2.1 and 3.8~kpc, respectively). The black dotted cone represents the distance traveled assuming the 90\% CI upper values for both SNR distance and NS proper motion. The green `x' is the 0.5--2.1~keV center of SNR emission and is a proxy for the ejecta's center of mass, and the red triangle is an estimate of the geometric center of the SNR. }}
\label{fig:NS_motion_G18.9}
\end{figure*}

\section{Proper Motion of the NS}
\label{sec:results}

We obtain proper motion measurements: 

\begin{align*}
\mu_{\alpha}\cos \delta & = -10.0 \pm 6.7 \rm{\ mas\ yr}^{-1} \\
\mu_{\delta} & = 22.6 \pm 6.8 \rm{\ mas\ yr}^{-1}
\end{align*}
corresponding to a total proper motion of 24.7 $\pm$ 6.8 mas yr$^{-1}$ at an angle of 336$^\circ$ $\pm$ 16$^\circ$ east of north. Using the two possible distance estimates of 2.1 $\pm$ 0.4 and 3.8 $\pm$0.4~kpc, this proper motion corresponds to transverse velocities of 246d$_{2.1}$ $\pm$ 78 km s$^{-1}$ and 445d$_{3.8}$ $\pm$ 129 km s$^{-1}$. Our measured direction of NS motion matches the morphology of the pulsar's elongated PWN \citep{tullmann10,bykov22}, which extends from the pulsar to the southeast, indicating that the pulsar is moving to the NW and will soon escape its PWN.

Finally, to obtain the most accurate NS kick velocity, we accounted for Galactic rotation and the peculiar motion of the sun. We used a solar distance to the Galactic center of 8.2~kpc, a solar peculiar velocity of v$_{ \odot}$ = (8.0, 12.4, and 7.7) km s$^{-1}$, and a Galactic rotation speed v$_{\rm LSR}$ of 236 km s$^{-1}$ with a flat rotation curve at these distances \citep{gawa19}. We found a correction of -0.21 (-0.60) mas yr$^{-1}$ in RA and -2.02 (-1.98) mas yr$^{-1}$ in Dec for a distance of 2.1 (3.8)~kpc. Thus, the transverse velocity of the NS in its local rest frame is either 264d$_{2.1}$ $\pm$ 78 km s$^{-1}$ or 474d$_{3.8}$ $\pm$ 129 km s$^{-1}$ depending on its distance.

\subsection{NS Velocity vs. Bulk Ejecta Motion}
\label{subsec:ejecta}

The 4400--6100~yr SNR age estimate of \cite{harrus04} assumes a 2~kpc distance, and the derived SNR age scales linearly with assumed distance. Thus, a distance of 2.1~kpc corresponds to an age of $\sim$5.5~kyr (4600--6400~yr), and a distance of 3.8~kpc corresponds to an age estimate of $\sim$10~kyr (8400--11600~yr.) If we back-evolve the NS motion of 24.7 $\pm$ 6.8 mas yr$^{-1}$ for 5.5~kyr and 10~kyr, we obtain total angular distances traveled of 2.25' $\pm$ 0.6' and 4.1' $\pm$ 1.1', respectively. 

Figure \ref{fig:NS_motion_G18.9} shows the back-evolved motion of the NS using both the 5.5~kyr and 10~kyr age estimates and accounting for the 1$\sigma$ uncertainty in kick direction. Currently, the NS is $\sim$9.5' away from G18.9's center of 0.5--2.1~keV X-ray emission (reflecting primarily shocked-heated ejecta) at an angle of 331$^\circ$ east of north, consistent to within 1$\sigma$ with the direction of NS motion.

 Based on the NS's calculated proper motion, its birth site is still multiple arcseconds NW from G18.9's center of 0.5--2.1~keV X-ray emission \citep{me17} and its geometric center---defined as the center of the circle or ellipse that approximately matches the SNR's X-ray profile. This remains true even if we assume the most extreme conditions---the highest age estimate (11,600~yr) and a velocity on the edge of the 90\% confidence interval (35.9 mas yr$^{-1}$)---corresponding to a total distance traveled of 7'. 

Taken together, these findings show that the NS in G18.9 has been kicked in a direction nearly opposite the bulk of ejecta. This finding is consistent with the conclusions of past papers that investigated the relationship between neutron star velocity and ejecta motion (e.g., \citealt{me17,katsuda18}). It provides further support for the dominating theory that NSs are accelerated due to a conservation-of-momentum-like process with the ejecta: i.e., the ``Gravitational Tugboat Mechanism'' of \citep{wongwathanarat13}.

\subsection{NS Velocity vs Ejecta Asymmetries}
\label{subsec:power-ratio}

\begin{figure*} 
\includegraphics[width=0.34\textwidth]{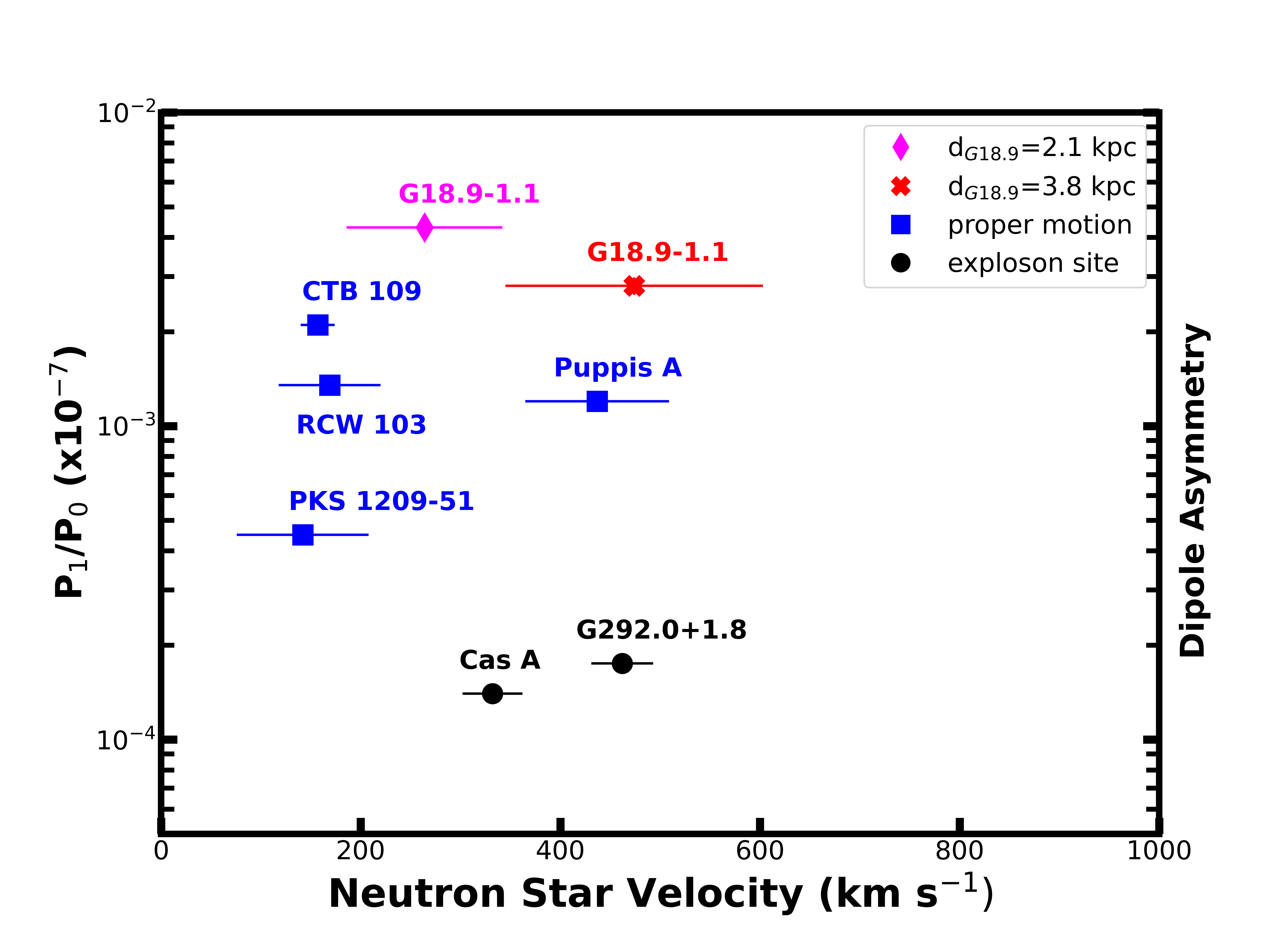}
\includegraphics[width=0.34\textwidth]{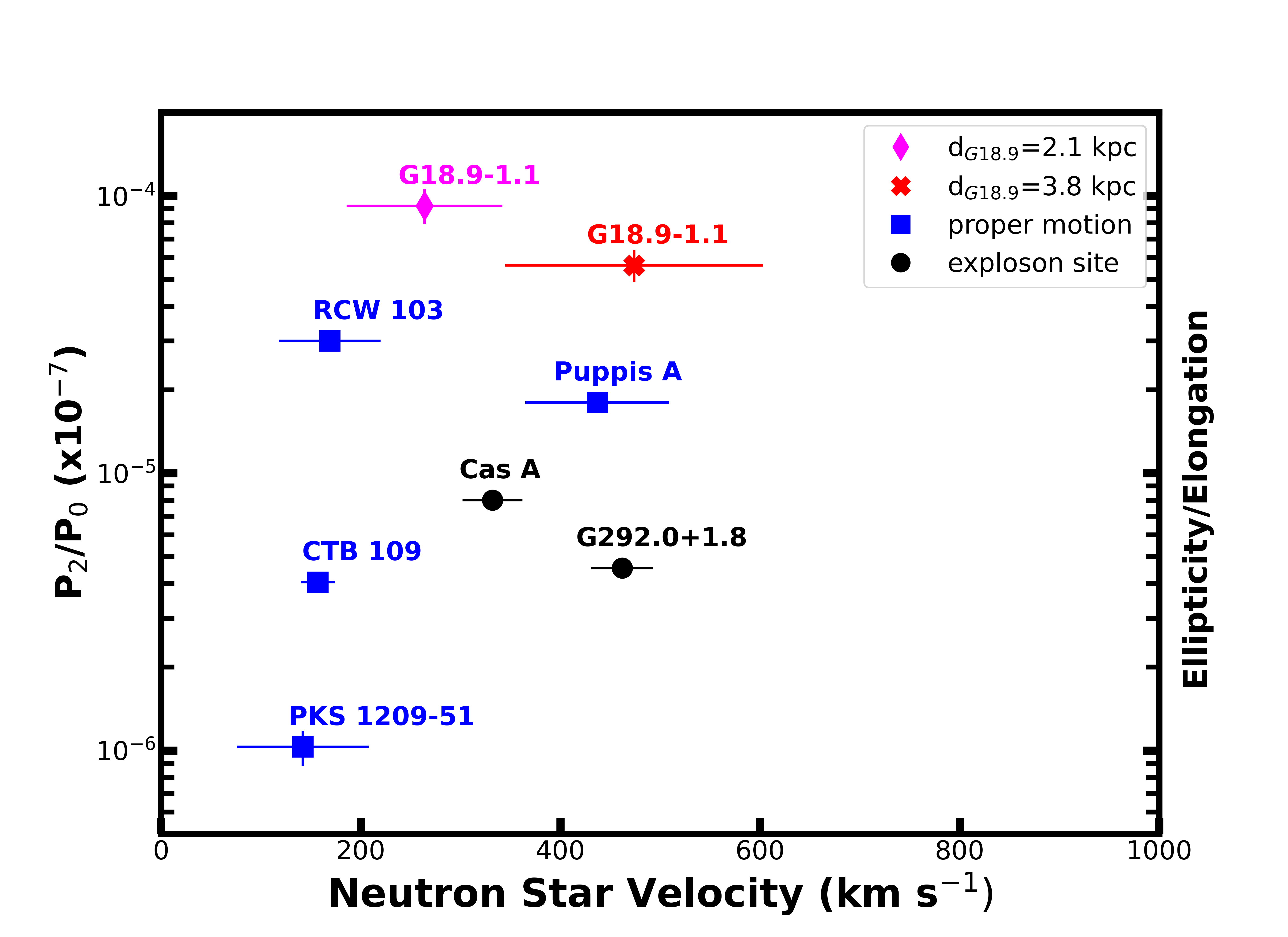}
\includegraphics[width=0.34\textwidth]{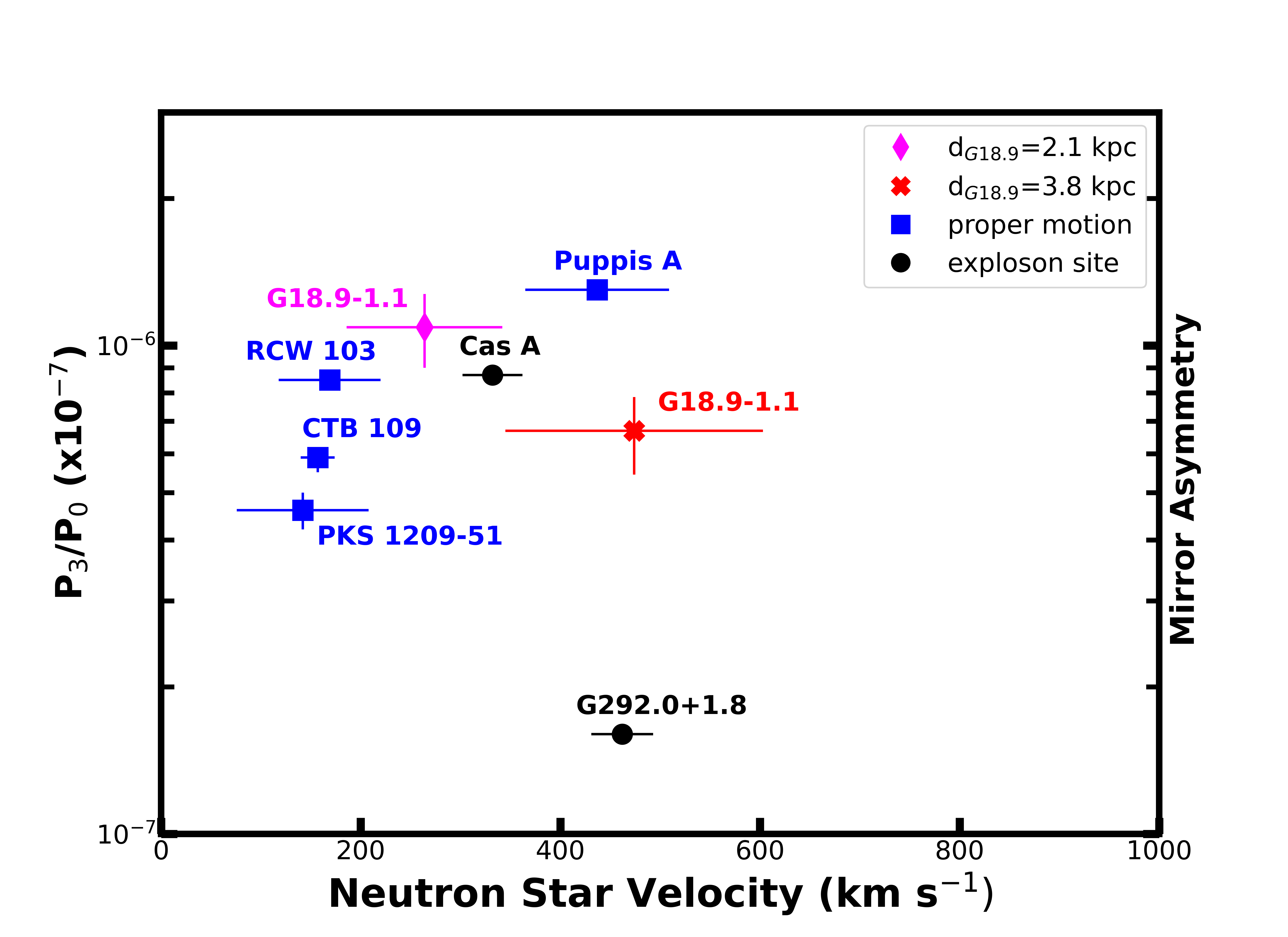}
\caption{The dipole (left), quadrupole (middle), and octupole (right) power ratios vs. neutron star velocities for the sample of SNRs with robust NS proper motions investigated in \cite{me17}, using the explosion site as the origin for analysis.  Blue points indicate NS velocities from direct proper motion measurements while black points indicate NS velocities obtained from back-evolved filament motion (Cas~A and G292.0$+$1.8). Circles indicate there is no evidence of SNR interaction with CSM/ISM, and squares indicate clear evidence of interaction. The magenta diamond (red `x') points assume a 2.1 (3.8)~kpc distance to and a 5.5 (10.0)~kyr age for SNR G18.9 and its associated NS.}
\label{fig:origin}
\end{figure*}

We then measured the 2D ejecta asymmetries of G18.9, using the power ratio method centered at both NS birth sites found via back-evolving the NS for 5.5~kyr and 10~kyr (associated with distances of 2.1 and 3.8~kpc, respectively). Figure~\ref{fig:origin}, showing the multipole moments vs. NS velocities of other SNRs that have well-constrained NS velocities, was taken from \cite{me17} and updated with the results from this paper. Each plot shows a different power ratio---reflecting the 2D dipole asymmetry (a proxy for bulk motion), ellipticity, and mirror asymmetry of the SNR---compared to the transverse velocity of the associated NS. 

The power ratios measured in SNRs depend on the observation angle, reaching maximum values if observed perpendicular to the SNR axis of symmetry and approaching zero if viewed along the symmetry axis. As the observation angle of SNRs can be difficult to estimate---requiring spatially-resolved Doppler analysis of the ejecta to create 3D SNR maps---there can be significant uncertainty when comparing power ratio values to parameters unaffected by viewing angle (e.g., explosion energy or magnetic field strength). However, as the component of the NS velocity projected onto the plane of the sky should be affected by the observation angle in the same manner as power ratios, the comparison between projected SNR ejecta moment distribution and transverse NS velocity does not depend on the symmetry axis orientation.

As shown, G18.9 exhibits generally stronger levels of asymmetry than other SNRs, although this difference is smaller when using the farther 3.8~kpc distance. Given that we have no other priors for favoring either SNR distance, we suggest that the farther distance of 3.8~kpc and larger NS velocity of $\sim$474~km s$^{-1}$ is slightly favored as the power ratios using this distance are closer to that of the other remnants analyzed in previous literature---particularly the more well-studied SNRs of Cassiopeia~A, G292.0$+$1.8, and Puppis~A. However, six remnants are a small sample and so this conclusion is tenuous.

\subsection{NS Velocity and Progenitor Mass}
\label{subsec:progenitor-mass}

Finally, we compared our NS velocity estimates with Figure 1 from \cite{burrows23}, who performed 3D simulations of CCSNe of various masses, including the effects on NS kick velocities in their simulation. Although we do not have any way of measuring the NS's radial motion, we can make a rough estimate by assuming its radial motion is equal in magnitude to its motion in each of the orthogonal directions. We multiply the measured 2D velocities by $\frac{\sqrt{3}}{\sqrt{2}}$ to obtain 3D velocity estimates of $\sim$323 or $\sim$580 km s$^{-1}$. Compared to their results, these velocities are most consistent with progenitor masses of 15-20M$_\odot$. This range is consistent with the finding that SNe G18.9's explosion energy is within the standard values for SN explosions \citep{harrus04, burrows21} and the progenitor estimate of $\sim$15--20M$_\odot$ made by \cite{bykov22} from ejecta mass estimates using {\it eROSITA} data.

\section{Conclusions}
\label{sec:conc}

In this paper, we have measured the proper motion of the NS CXOU J182913.1--125113 in the SNR G18.9$-$1.1 for the first time using 15~years of {\it Chandra} data. We accounted for PSF effects to obtain the most accurate point source positions, corrected the astrometry of both epochs' observations using registration Gaia sources, and accounted for Galactic rotation and the peculiar motion of sun. We measured a proper motion of ($\mu_{\alpha}\cos \delta$, $\mu_\delta$) = (-10.0 $\pm$ 6.7, 22.6 $\pm$ 6.8) for a total proper motion $\mu$ = 24.7 $\pm$ 6.8 mas yr$^{-1}$ at an angle of 336$^\circ$ $\pm$ 16$^\circ$ east of north. Using the literature-reported distance estimates of 2.1 $\pm$ 0.4 and 3.8 $\pm$ 0.4~kpc, this proper motion corresponds to transverse velocities of 264d$_{2.1}$ $\pm$ 78 km s$^{-1}$ and 474d$_{3.8}$ $\pm$ 129 km s$^{-1}$, respectively. Based on multipole analysis using the power ratio method, we {\it slightly} favor the larger SNR distance and correspondingly faster NS velocity. As this farther distance corresponds to an older SNR age, this conclusion also favors the lower-temperature, CIE spectral fits of \cite{bykov22} over their higher-temperature, NEI fits.

Both of our velocity estimates fall within typical NS velocities of a few hundred km s$^{-1}$, well supported by various other observations and simulations. In particular, the 90\% confidence intervals of our estimates are inconsistent with the geometric velocity estimate of 700--960 km s$^{-1}$, providing further evidence against using the geometric center of a CCSNR as the NS's birth site or SNR explosion site. Additionally, the nearly-180$^\circ$ separation between the NS kick and the direction of bulk ejecta motion provides further support for the scenario where the NS is kicked due to conservation-of-momentum-like forces between it and the SN ejecta: a theory which has had growing theoretical and observation evidence over the past decade (see e.g., \citealt{wongwathanarat13,me17,katsuda18,me20a,burrows23}.

Future X-ray telescopes with high spatial resolution are vital for continuing to constrain the velocities of young NSs. The {\it Chandra} archive contains numerous observations of NSs from going back to 1999, so observations of NSs in the 2030s and 2040s can provide 20--40 year baselines corresponding to total NS motions of $\sim$0.5--10''. Thus, the proposed probe mission {\it AXIS} and the potential future mission {\it Lynx} would easily be able to measure NS motions. In particular, X-ray telescopes with small off-axis PSFs will enable more robust detections of registration point sources with which to better correct image astrometry. Until that time, Chandra remains the only instrument capable of measuring the proper motions of neutron stars.

\acknowledgments
\noindent \textbf{Acknowledgements}

Tyler Holland-Ashford’s research was supported by an appointment to the NASA Postdoctoral Program at the NASA Goddard Space Flight Center, administered by Oak Ridge Associated Universities under contract with NASA. Dr. Pat Slane acknowledges support from NASA Contract NAS8-03060. X.L. is supported by a GRF grant of the Hong Kong Government under HKU 17304524. Support for this work was provided by the National Aeronautics and Space Administration through Chandra Award Number GO3-24054Z issued by the Chandra X-ray Center, which is operated by the Smithsonian Astrophysical Observatory for and on behalf of the National Aeronautics Space Administration under contract NAS8-03060. The authors thank Dr. Tom Aldcroft for helpful discussions on Chandra astrometry.

This paper employs a list of Chandra datasets, obtained by the Chandra X-ray Observatory, contained in~\dataset[Chandra Data Collection (CDC)  366]{https://doi.org/10.25574/cdc.366}. This research has made use of data obtained from the {\it Chandra Data Archive}, software provided by the {\it Chandra X-ray Center} (CXC) in the application packages CIAO (4.13, \citealt{fruscione06}) and Sherpa (4.13.0, \citealt{sherpa,freeman01}), and the Python library SciPy \citep{scipy20}. This work made use of data from the European Space Agency (ESA) mission {\it Gaia} (\url{https://www.cosmos.esa.int/gaia}), processed by the {\it Gaia} Data Processing and Analysis Consortium (DPAC, \url{https://www.cosmos.esa.int/web/gaia/dpac/consortium}). Funding for the DPAC has been provided by national institutions, in particular the institutions participating in the {\it Gaia} Multilateral Agreement.

\bibliography{G18pt9_NS_Vel}

\begin{thebibliography}{}
\expandafter\ifx\csname natexlab\endcsname\relax\def\natexlab#1{#1}\fi

\bibitem[{{Babusiaux} {et~al.}(2023){Babusiaux}, {Fabricius}, {Khanna}, {Muraveva}, {Reyl{\'e}}, {Spoto}, {Vallenari}, {Luri}, {Arenou}, {{\'A}lvarez}, {Anders}, {Antoja}, {Balbinot}, {Barache}, {Bauchet}, {Bossini}, {Busonero}, {Cantat-Gaudin}, {Carrasco}, {Dafonte}, {Diakit{\'e}}, {Figueras}, {Garcia-Gutierrez}, {Garofalo}, {Helmi}, {Jim{\'e}nez-Arranz}, {Jordi}, {Kervella}, {Kostrzewa-Rutkowska}, {Leclerc}, {Licata}, {Manteiga}, {Masip}, {Mongui{\'o}}, {Ramos}, {Robichon}, {Robin}, {Romero-G{\'o}mez}, {S{\'a}ez}, {Santove{\~n}a}, {Spina}, {Torralba Elipe}, \& {Weiler}}]{babusiaux23}
{Babusiaux}, C., {Fabricius}, C., {Khanna}, S., {et~al.} 2023, \aap, 674, A32

\bibitem[{{Banovetz} {et~al.}(2021){Banovetz}, {Milisavljevic}, {Sravan}, {Fesen}, {Patnaude}, {Plucinsky}, {Blair}, {Weil}, {Morse}, {Margutti}, \& {Drout}}]{banovetz21}
{Banovetz}, J., {Milisavljevic}, D., {Sravan}, N., {et~al.} 2021, \apj, 912, 33

\bibitem[{Burke {et~al.}(2021)Burke, Laurino, wmclaugh, Marie-Terrell, dtnguyen2, Günther, Siemiginowska, Budynkiewicz, Cheer, Aldcroft, Deil, Sipőcz, Buchner, Donath, Laginja, Leinweber, nplee, \& Todd}]{sherpa}
Burke, D., Laurino, O., wmclaugh, {et~al.} 2021, sherpa/sherpa: Sherpa 4.13.0, doi:10.5281/zenodo.4428938

\bibitem[{{Burrows} \& {Vartanyan}(2021)}]{burrows21}
{Burrows}, A., \& {Vartanyan}, D. 2021, \nat, 589, 29

\bibitem[{{Burrows} {et~al.}(2023){Burrows}, {Wang}, {Vartanyan}, \& {Coleman}}]{burrows23}
{Burrows}, A., {Wang}, T., {Vartanyan}, D., \& {Coleman}, M. S.~B. 2023, arXiv e-prints, arXiv:2311.12109

\bibitem[{{Bykov} {et~al.}(2022){Bykov}, {Uvarov}, {Churazov}, {Gilfanov}, \& {Medvedev}}]{bykov22}
{Bykov}, A.~M., {Uvarov}, Y.~A., {Churazov}, E.~M., {Gilfanov}, M.~R., \& {Medvedev}, P.~S. 2022, \aap, 661, A19

\bibitem[{{Coleman} \& {Burrows}(2022)}]{coleman22}
{Coleman}, M. S.~B., \& {Burrows}, A. 2022, \mnras, 517, 3938

\bibitem[{{De Luca}(2017)}]{deluca17}
{De Luca}, A. 2017, in Journal of Physics Conference Series, Vol. 932, Journal of Physics Conference Series (IOP), 012006

\bibitem[{{de Vries} {et~al.}(2021){de Vries}, {Romani}, {Kargaltsev}, {Pavlov}, {Posselt}, {Slane}, {Bucciantini}, {Ng}, \& {Klingler}}]{devries21}
{de Vries}, M., {Romani}, R.~W., {Kargaltsev}, O., {et~al.} 2021, \apj, 908, 50

\bibitem[{{Dzib} \& {Rodr{\'\i}guez}(2021)}]{dzib21}
{Dzib}, S.~A., \& {Rodr{\'\i}guez}, L.~F. 2021, \apj, 923, 228

\bibitem[{{Fesen} {et~al.}(2006){Fesen}, {Hammell}, {Morse}, {Chevalier}, {Borkowski}, {Dopita}, {Gerardy}, {Lawrence}, {Raymond}, \& {van den Bergh}}]{fesen06}
{Fesen}, R.~A., {Hammell}, M.~C., {Morse}, J., {et~al.} 2006, \apj, 645, 283

\bibitem[{{Frail} {et~al.}(1996){Frail}, {Giacani}, {Goss}, \& {Dubner}}]{frail96}
{Frail}, D.~A., {Giacani}, E.~B., {Goss}, W.~M., \& {Dubner}, G. 1996, \apjl, 464, L165

\bibitem[{{Freeman} {et~al.}(2001){Freeman}, {Doe}, \& {Siemiginowska}}]{freeman01}
{Freeman}, P., {Doe}, S., \& {Siemiginowska}, A. 2001, in Society of Photo-Optical Instrumentation Engineers (SPIE) Conference Series, Vol. 4477, Astronomical Data Analysis, ed. J.-L. {Starck} \& F.~D. {Murtagh}, 76--87

\bibitem[{{Fruscione} {et~al.}(2006){Fruscione}, {McDowell}, {Allen}, {Brickhouse}, {Burke}, {Davis}, {Durham}, {Elvis}, {Galle}, {Harris}, {Huenemoerder}, {Houck}, {Ishibashi}, {Karovska}, {Nicastro}, {Noble}, {Nowak}, {Primini}, {Siemiginowska}, {Smith}, \& {Wise}}]{fruscione06}
{Fruscione}, A., {McDowell}, J.~C., {Allen}, G.~E., {et~al.} 2006, in Society of Photo-Optical Instrumentation Engineers (SPIE) Conference Series, Vol. 6270, Society of Photo-Optical Instrumentation Engineers (SPIE) Conference Series, ed. D.~R. {Silva} \& R.~E. {Doxsey}, 62701V

\bibitem[{{Fryer} \& {Kusenko}(2006)}]{fryer06}
{Fryer}, C.~L., \& {Kusenko}, A. 2006, \apjs, 163, 335

\bibitem[{{Fuerst} {et~al.}(1997){Fuerst}, {Reich}, \& {Aschenbach}}]{fuerst97}
{Fuerst}, E., {Reich}, W., \& {Aschenbach}, B. 1997, \aap, 319, 655

\bibitem[{{Gaia Collaboration} {et~al.}(2016){Gaia Collaboration}, {Prusti}, {de Bruijne}, {Brown}, {Vallenari}, {Babusiaux}, {Bailer-Jones}, {Bastian}, {Biermann}, {Evans}, {Eyer}, {Jansen}, {Jordi}, {Klioner}, {Lammers}, {Lindegren}, {Luri}, {Mignard}, {Milligan}, {Panem}, {Poinsignon}, {Pourbaix}, {Randich}, {Sarri}, {Sartoretti}, {Siddiqui}, {Soubiran}, {Valette}, {van Leeuwen}, {Walton}, {Aerts}, {Arenou}, {Cropper}, {Drimmel}, {H{\o}g}, {Katz}, {Lattanzi}, {O'Mullane}, {Grebel}, {Holland}, {Huc}, {Passot}, {Bramante}, {Cacciari}, {Casta{\~n}eda}, {Chaoul}, {Cheek}, {De Angeli}, {Fabricius}, {Guerra}, {Hern{\'a}ndez}, {Jean-Antoine-Piccolo}, {Masana}, {Messineo}, {Mowlavi}, {Nienartowicz}, {Ord{\'o}{\~n}ez-Blanco}, {Panuzzo}, {Portell}, {Richards}, {Riello}, {Seabroke}, {Tanga}, {Th{\'e}venin}, {Torra}, {Els}, {Gracia-Abril}, {Comoretto}, {Garcia-Reinaldos}, {Lock}, {Mercier}, {Altmann}, {Andrae}, {Astraatmadja}, {Bellas-Velidis}, {Benson}, {Berthier}, {Blomme}, {Busso}, {Carry}, {Cellino}, {Clementini},
  {Cowell}, {Creevey}, {Cuypers}, {Davidson}, {De Ridder}, {de Torres}, {Delchambre}, {Dell'Oro}, {Ducourant}, {Fr{\'e}mat}, {Garc{\'\i}a-Torres}, {Gosset}, {Halbwachs}, {Hambly}, {Harrison}, {Hauser}, {Hestroffer}, {Hodgkin}, {Huckle}, {Hutton}, {Jasniewicz}, {Jordan}, {Kontizas}, {Korn}, {Lanzafame}, {Manteiga}, {Moitinho}, {Muinonen}, {Osinde}, {Pancino}, {Pauwels}, {Petit}, {Recio-Blanco}, {Robin}, {Sarro}, {Siopis}, {Smith}, {Smith}, {Sozzetti}, {Thuillot}, {van Reeven}, {Viala}, {Abbas}, {Abreu Aramburu}, {Accart}, {Aguado}, {Allan}, {Allasia}, {Altavilla}, {{\'A}lvarez}, {Alves}, {Anderson}, {Andrei}, {Anglada Varela}, {Antiche}, {Antoja}, {Ant{\'o}n}, {Arcay}, {Atzei}, {Ayache}, {Bach}, {Baker}, {Balaguer-N{\'u}{\~n}ez}, {Barache}, {Barata}, {Barbier}, {Barblan}, {Baroni}, {Barrado y Navascu{\'e}s}, {Barros}, {Barstow}, {Becciani}, {Bellazzini}, {Bellei}, {Bello Garc{\'\i}a}, {Belokurov}, {Bendjoya}, {Berihuete}, {Bianchi}, {Bienaym{\'e}}, {Billebaud}, {Blagorodnova}, {Blanco-Cuaresma}, {Boch},
  {Bombrun}, {Borrachero}, {Bouquillon}, {Bourda}, {Bouy}, {Bragaglia}, {Breddels}, {Brouillet}, {Br{\"u}semeister}, {Bucciarelli}, {Budnik}, {Burgess}, {Burgon}, {Burlacu}, {Busonero}, {Buzzi}, {Caffau}, {Cambras}, {Campbell}, {Cancelliere}, {Cantat-Gaudin}, {Carlucci}, {Carrasco}, {Castellani}, {Charlot}, {Charnas}, {Charvet}, {Chassat}, {Chiavassa}, {Clotet}, {Cocozza}, {Collins}, {Collins}, {Costigan}, {Crifo}, {Cross}, {Crosta}, {Crowley}, {Dafonte}, {Damerdji}, {Dapergolas}, {David}, {David}, {De Cat}, {de Felice}, {de Laverny}, {De Luise}, {De March}, {de Martino}, {de Souza}, {Debosscher}, {del Pozo}, {Delbo}, {Delgado}, {Delgado}, {di Marco}, {Di Matteo}, {Diakite}, {Distefano}, {Dolding}, {Dos Anjos}, {Drazinos}, {Dur{\'a}n}, {Dzigan}, {Ecale}, {Edvardsson}, {Enke}, {Erdmann}, {Escolar}, {Espina}, {Evans}, {Eynard Bontemps}, {Fabre}, {Fabrizio}, {Faigler}, {Falc{\~a}o}, {Farr{\`a}s Casas}, {Faye}, {Federici}, {Fedorets}, {Fern{\'a}ndez-Hern{\'a}ndez}, {Fernique}, {Fienga}, {Figueras}, {Filippi},
  {Findeisen}, {Fonti}, {Fouesneau}, {Fraile}, {Fraser}, {Fuchs}, {Furnell}, {Gai}, {Galleti}, {Galluccio}, {Garabato}, {Garc{\'\i}a-Sedano}, {Gar{\'e}}, {Garofalo}, {Garralda}, {Gavras}, {Gerssen}, {Geyer}, {Gilmore}, {Girona}, {Giuffrida}, {Gomes}, {Gonz{\'a}lez-Marcos}, {Gonz{\'a}lez-N{\'u}{\~n}ez}, {Gonz{\'a}lez-Vidal}, {Granvik}, {Guerrier}, {Guillout}, {Guiraud}, {G{\'u}rpide}, {Guti{\'e}rrez-S{\'a}nchez}, {Guy}, {Haigron}, {Hatzidimitriou}, {Haywood}, {Heiter}, {Helmi}, {Hobbs}, {Hofmann}, {Holl}, {Holland}, {Hunt}, {Hypki}, {Icardi}, {Irwin}, {Jevardat de Fombelle}, {Jofr{\'e}}, {Jonker}, {Jorissen}, {Julbe}, {Karampelas}, {Kochoska}, {Kohley}, {Kolenberg}, {Kontizas}, {Koposov}, {Kordopatis}, {Koubsky}, {Kowalczyk}, {Krone-Martins}, {Kudryashova}, {Kull}, {Bachchan}, {Lacoste-Seris}, {Lanza}, {Lavigne}, {Le Poncin-Lafitte}, {Lebreton}, {Lebzelter}, {Leccia}, {Leclerc}, {Lecoeur-Taibi}, {Lemaitre}, {Lenhardt}, {Leroux}, {Liao}, {Licata}, {Lindstr{\o}m}, {Lister}, {Livanou}, {Lobel}, {L{\"o}ffler},
  {L{\'o}pez}, {Lopez-Lozano}, {Lorenz}, {Loureiro}, {MacDonald}, {Magalh{\~a}es Fernandes}, {Managau}, {Mann}, {Mantelet}, {Marchal}, {Marchant}, {Marconi}, {Marie}, {Marinoni}, {Marrese}, {Marschalk{\'o}}, {Marshall}, {Mart{\'\i}n-Fleitas}, {Martino}, {Mary}, {Matijevi{\v{c}}}, {Mazeh}, {McMillan}, {Messina}, {Mestre}, {Michalik}, {Millar}, {Miranda}, {Molina}, {Molinaro}, {Molinaro}, {Moln{\'a}r}, {Moniez}, {Montegriffo}, {Monteiro}, {Mor}, {Mora}, {Morbidelli}, {Morel}, {Morgenthaler}, {Morley}, {Morris}, {Mulone}, {Muraveva}, {Musella}, {Narbonne}, {Nelemans}, {Nicastro}, {Noval}, {Ord{\'e}novic}, {Ordieres-Mer{\'e}}, {Osborne}, {Pagani}, {Pagano}, {Pailler}, {Palacin}, {Palaversa}, {Parsons}, {Paulsen}, {Pecoraro}, {Pedrosa}, {Pentik{\"a}inen}, {Pereira}, {Pichon}, {Piersimoni}, {Pineau}, {Plachy}, {Plum}, {Poujoulet}, {Pr{\v{s}}a}, {Pulone}, {Ragaini}, {Rago}, {Rambaux}, {Ramos-Lerate}, {Ranalli}, {Rauw}, {Read}, {Regibo}, {Renk}, {Reyl{\'e}}, {Ribeiro}, {Rimoldini}, {Ripepi}, {Riva}, {Rixon},
  {Roelens}, {Romero-G{\'o}mez}, {Rowell}, {Royer}, {Rudolph}, {Ruiz-Dern}, {Sadowski}, {Sagrist{\`a} Sell{\'e}s}, {Sahlmann}, {Salgado}, {Salguero}, {Sarasso}, {Savietto}, {Schnorhk}, {Schultheis}, {Sciacca}, {Segol}, {Segovia}, {Segransan}, {Serpell}, {Shih}, {Smareglia}, {Smart}, {Smith}, {Solano}, {Solitro}, {Sordo}, {Soria Nieto}, {Souchay}, {Spagna}, {Spoto}, {Stampa}, {Steele}, {Steidelm{\"u}ller}, {Stephenson}, {Stoev}, {Suess}, {S{\"u}veges}, {Surdej}, {Szabados}, {Szegedi-Elek}, {Tapiador}, {Taris}, {Tauran}, {Taylor}, {Teixeira}, {Terrett}, {Tingley}, {Trager}, {Turon}, {Ulla}, {Utrilla}, {Valentini}, {van Elteren}, {Van Hemelryck}, {van Leeuwen}, {Varadi}, {Vecchiato}, {Veljanoski}, {Via}, {Vicente}, {Vogt}, {Voss}, {Votruba}, {Voutsinas}, {Walmsley}, {Weiler}, {Weingrill}, {Werner}, {Wevers}, {Whitehead}, {Wyrzykowski}, {Yoldas}, {{\v{Z}}erjal}, {Zucker}, {Zurbach}, {Zwitter}, {Alecu}, {Allen}, {Allende Prieto}, {Amorim}, {Anglada-Escud{\'e}}, {Arsenijevic}, {Azaz}, {Balm}, {Beck}, {Bernstein},
  {Bigot}, {Bijaoui}, {Blasco}, {Bonfigli}, {Bono}, {Boudreault}, {Bressan}, {Brown}, {Brunet}, {Bunclark}, {Buonanno}, {Butkevich}, {Carret}, {Carrion}, {Chemin}, {Ch{\'e}reau}, {Corcione}, {Darmigny}, {de Boer}, {de Teodoro}, {de Zeeuw}, {Delle Luche}, {Domingues}, {Dubath}, {Fodor}, {Fr{\'e}zouls}, {Fries}, {Fustes}, {Fyfe}, {Gallardo}, {Gallegos}, {Gardiol}, {Gebran}, {Gomboc}, {G{\'o}mez}, {Grux}, {Gueguen}, {Heyrovsky}, {Hoar}, {Iannicola}, {Isasi Parache}, {Janotto}, {Joliet}, {Jonckheere}, {Keil}, {Kim}, {Klagyivik}, {Klar}, {Knude}, {Kochukhov}, {Kolka}, {Kos}, {Kutka}, {Lainey}, {LeBouquin}, {Liu}, {Loreggia}, {Makarov}, {Marseille}, {Martayan}, {Martinez-Rubi}, {Massart}, {Meynadier}, {Mignot}, {Munari}, {Nguyen}, {Nordlander}, {Ocvirk}, {O'Flaherty}, {Olias Sanz}, {Ortiz}, {Osorio}, {Oszkiewicz}, {Ouzounis}, {Palmer}, {Park}, {Pasquato}, {Peltzer}, {Peralta}, {P{\'e}turaud}, {Pieniluoma}, {Pigozzi}, {Poels}, {Prat}, {Prod'homme}, {Raison}, {Rebordao}, {Risquez}, {Rocca-Volmerange}, {Rosen},
  {Ruiz-Fuertes}, {Russo}, {Sembay}, {Serraller Vizcaino}, {Short}, {Siebert}, {Silva}, {Sinachopoulos}, {Slezak}, {Soffel}, {Sosnowska}, {Strai{\v{z}}ys}, {ter Linden}, {Terrell}, {Theil}, {Tiede}, {Troisi}, {Tsalmantza}, {Tur}, {Vaccari}, {Vachier}, {Valles}, {Van Hamme}, {Veltz}, {Virtanen}, {Wallut}, {Wichmann}, {Wilkinson}, {Ziaeepour}, \& {Zschocke}}]{gaia16b}
{Gaia Collaboration}, {Prusti}, T., {de Bruijne}, J.~H.~J., {et~al.} 2016, \aap, 595, A1

\bibitem[{{Gaia Collaboration} {et~al.}(2023){Gaia Collaboration}, {Vallenari}, {Brown}, {Prusti}, {de Bruijne}, {Arenou}, {Babusiaux}, {Biermann}, {Creevey}, {Ducourant}, {Evans}, {Eyer}, {Guerra}, {Hutton}, {Jordi}, {Klioner}, {Lammers}, {Lindegren}, {Luri}, {Mignard}, {Panem}, {Pourbaix}, {Randich}, {Sartoretti}, {Soubiran}, {Tanga}, {Walton}, {Bailer-Jones}, {Bastian}, {Drimmel}, {Jansen}, {Katz}, {Lattanzi}, {van Leeuwen}, {Bakker}, {Cacciari}, {Casta{\~n}eda}, {De Angeli}, {Fabricius}, {Fouesneau}, {Fr{\'e}mat}, {Galluccio}, {Guerrier}, {Heiter}, {Masana}, {Messineo}, {Mowlavi}, {Nicolas}, {Nienartowicz}, {Pailler}, {Panuzzo}, {Riclet}, {Roux}, {Seabroke}, {Sordo}, {Th{\'e}venin}, {Gracia-Abril}, {Portell}, {Teyssier}, {Altmann}, {Andrae}, {Audard}, {Bellas-Velidis}, {Benson}, {Berthier}, {Blomme}, {Burgess}, {Busonero}, {Busso}, {C{\'a}novas}, {Carry}, {Cellino}, {Cheek}, {Clementini}, {Damerdji}, {Davidson}, {de Teodoro}, {Nu{\~n}ez Campos}, {Delchambre}, {Dell'Oro}, {Esquej},
  {Fern{\'a}ndez-Hern{\'a}ndez}, {Fraile}, {Garabato}, {Garc{\'\i}a-Lario}, {Gosset}, {Haigron}, {Halbwachs}, {Hambly}, {Harrison}, {Hern{\'a}ndez}, {Hestroffer}, {Hodgkin}, {Holl}, {Jan{\ss}en}, {Jevardat de Fombelle}, {Jordan}, {Krone-Martins}, {Lanzafame}, {L{\"o}ffler}, {Marchal}, {Marrese}, {Moitinho}, {Muinonen}, {Osborne}, {Pancino}, {Pauwels}, {Recio-Blanco}, {Reyl{\'e}}, {Riello}, {Rimoldini}, {Roegiers}, {Rybizki}, {Sarro}, {Siopis}, {Smith}, {Sozzetti}, {Utrilla}, {van Leeuwen}, {Abbas}, {{\'A}brah{\'a}m}, {Abreu Aramburu}, {Aerts}, {Aguado}, {Ajaj}, {Aldea-Montero}, {Altavilla}, {{\'A}lvarez}, {Alves}, {Anders}, {Anderson}, {Anglada Varela}, {Antoja}, {Baines}, {Baker}, {Balaguer-N{\'u}{\~n}ez}, {Balbinot}, {Balog}, {Barache}, {Barbato}, {Barros}, {Barstow}, {Bartolom{\'e}}, {Bassilana}, {Bauchet}, {Becciani}, {Bellazzini}, {Berihuete}, {Bernet}, {Bertone}, {Bianchi}, {Binnenfeld}, {Blanco-Cuaresma}, {Blazere}, {Boch}, {Bombrun}, {Bossini}, {Bouquillon}, {Bragaglia}, {Bramante}, {Breedt},
  {Bressan}, {Brouillet}, {Brugaletta}, {Bucciarelli}, {Burlacu}, {Butkevich}, {Buzzi}, {Caffau}, {Cancelliere}, {Cantat-Gaudin}, {Carballo}, {Carlucci}, {Carnerero}, {Carrasco}, {Casamiquela}, {Castellani}, {Castro-Ginard}, {Chaoul}, {Charlot}, {Chemin}, {Chiaramida}, {Chiavassa}, {Chornay}, {Comoretto}, {Contursi}, {Cooper}, {Cornez}, {Cowell}, {Crifo}, {Cropper}, {Crosta}, {Crowley}, {Dafonte}, {Dapergolas}, {David}, {David}, {de Laverny}, {De Luise}, {De March}, {De Ridder}, {de Souza}, {de Torres}, {del Peloso}, {del Pozo}, {Delbo}, {Delgado}, {Delisle}, {Demouchy}, {Dharmawardena}, {Di Matteo}, {Diakite}, {Diener}, {Distefano}, {Dolding}, {Edvardsson}, {Enke}, {Fabre}, {Fabrizio}, {Faigler}, {Fedorets}, {Fernique}, {Fienga}, {Figueras}, {Fournier}, {Fouron}, {Fragkoudi}, {Gai}, {Garcia-Gutierrez}, {Garcia-Reinaldos}, {Garc{\'\i}a-Torres}, {Garofalo}, {Gavel}, {Gavras}, {Gerlach}, {Geyer}, {Giacobbe}, {Gilmore}, {Girona}, {Giuffrida}, {Gomel}, {Gomez}, {Gonz{\'a}lez-N{\'u}{\~n}ez},
  {Gonz{\'a}lez-Santamar{\'\i}a}, {Gonz{\'a}lez-Vidal}, {Granvik}, {Guillout}, {Guiraud}, {Guti{\'e}rrez-S{\'a}nchez}, {Guy}, {Hatzidimitriou}, {Hauser}, {Haywood}, {Helmer}, {Helmi}, {Sarmiento}, {Hidalgo}, {Hilger}, {H{\l}adczuk}, {Hobbs}, {Holland}, {Huckle}, {Jardine}, {Jasniewicz}, {Jean-Antoine Piccolo}, {Jim{\'e}nez-Arranz}, {Jorissen}, {Juaristi Campillo}, {Julbe}, {Karbevska}, {Kervella}, {Khanna}, {Kontizas}, {Kordopatis}, {Korn}, {K{\'o}sp{\'a}l}, {Kostrzewa-Rutkowska}, {Kruszy{\'n}ska}, {Kun}, {Laizeau}, {Lambert}, {Lanza}, {Lasne}, {Le Campion}, {Lebreton}, {Lebzelter}, {Leccia}, {Leclerc}, {Lecoeur-Taibi}, {Liao}, {Licata}, {Lindstr{\o}m}, {Lister}, {Livanou}, {Lobel}, {Lorca}, {Loup}, {Madrero Pardo}, {Magdaleno Romeo}, {Managau}, {Mann}, {Manteiga}, {Marchant}, {Marconi}, {Marcos}, {Marcos Santos}, {Mar{\'\i}n Pina}, {Marinoni}, {Marocco}, {Marshall}, {Martin Polo}, {Mart{\'\i}n-Fleitas}, {Marton}, {Mary}, {Masip}, {Massari}, {Mastrobuono-Battisti}, {Mazeh}, {McMillan}, {Messina}, {Michalik},
  {Millar}, {Mints}, {Molina}, {Molinaro}, {Moln{\'a}r}, {Monari}, {Mongui{\'o}}, {Montegriffo}, {Montero}, {Mor}, {Mora}, {Morbidelli}, {Morel}, {Morris}, {Muraveva}, {Murphy}, {Musella}, {Nagy}, {Noval}, {Oca{\~n}a}, {Ogden}, {Ordenovic}, {Osinde}, {Pagani}, {Pagano}, {Palaversa}, {Palicio}, {Pallas-Quintela}, {Panahi}, {Payne-Wardenaar}, {Pe{\~n}alosa Esteller}, {Penttil{\"a}}, {Pichon}, {Piersimoni}, {Pineau}, {Plachy}, {Plum}, {Poggio}, {Pr{\v{s}}a}, {Pulone}, {Racero}, {Ragaini}, {Rainer}, {Raiteri}, {Rambaux}, {Ramos}, {Ramos-Lerate}, {Re Fiorentin}, {Regibo}, {Richards}, {Rios Diaz}, {Ripepi}, {Riva}, {Rix}, {Rixon}, {Robichon}, {Robin}, {Robin}, {Roelens}, {Rogues}, {Rohrbasser}, {Romero-G{\'o}mez}, {Rowell}, {Royer}, {Ruz Mieres}, {Rybicki}, {Sadowski}, {S{\'a}ez N{\'u}{\~n}ez}, {Sagrist{\`a} Sell{\'e}s}, {Sahlmann}, {Salguero}, {Samaras}, {Sanchez Gimenez}, {Sanna}, {Santove{\~n}a}, {Sarasso}, {Schultheis}, {Sciacca}, {Segol}, {Segovia}, {S{\'e}gransan}, {Semeux}, {Shahaf}, {Siddiqui}, {Siebert},
  {Siltala}, {Silvelo}, {Slezak}, {Slezak}, {Smart}, {Snaith}, {Solano}, {Solitro}, {Souami}, {Souchay}, {Spagna}, {Spina}, {Spoto}, {Steele}, {Steidelm{\"u}ller}, {Stephenson}, {S{\"u}veges}, {Surdej}, {Szabados}, {Szegedi-Elek}, {Taris}, {Taylor}, {Teixeira}, {Tolomei}, {Tonello}, {Torra}, {Torra}, {Torralba Elipe}, {Trabucchi}, {Tsounis}, {Turon}, {Ulla}, {Unger}, {Vaillant}, {van Dillen}, {van Reeven}, {Vanel}, {Vecchiato}, {Viala}, {Vicente}, {Voutsinas}, {Weiler}, {Wevers}, {Wyrzykowski}, {Yoldas}, {Yvard}, {Zhao}, {Zorec}, {Zucker}, \& {Zwitter}}]{gaia23j}
{Gaia Collaboration}, {Vallenari}, A., {Brown}, A.~G.~A., {et~al.} 2023, \aap, 674, A1

\bibitem[{{Halpern} \& {Gotthelf}(2015)}]{halpern15}
{Halpern}, J.~P., \& {Gotthelf}, E.~V. 2015, \apj, 812, 61

\bibitem[{{Harrus} {et~al.}(2004){Harrus}, {Slane}, {Hughes}, \& {Plucinsky}}]{harrus04}
{Harrus}, I.~M., {Slane}, P.~O., {Hughes}, J.~P., \& {Plucinsky}, P.~P. 2004, \apj, 603, 152

\bibitem[{{Ho} {et~al.}(2020){Ho}, {Guillot}, {Saz Parkinson}, {Limyansky}, {Ng}, {Bejger}, {Espinoza}, {Haskell}, {Jaisawal}, \& {Malacaria}}]{ho20}
{Ho}, W. C.~G., {Guillot}, S., {Saz Parkinson}, P.~M., {et~al.} 2020, \mnras, 498, 4396

\bibitem[{{Hobbs} {et~al.}(2005){Hobbs}, {Lorimer}, {Lyne}, \& {Kramer}}]{hobbs05}
{Hobbs}, G., {Lorimer}, D.~R., {Lyne}, A.~G., \& {Kramer}, M. 2005, \mnras, 360, 974

\bibitem[{{Holland-Ashford} {et~al.}(2020){Holland-Ashford}, {Lopez}, \& {Auchettl}}]{me20a}
{Holland-Ashford}, T., {Lopez}, L.~A., \& {Auchettl}, K. 2020, \apj, 889, 144

\bibitem[{{Holland-Ashford} {et~al.}(2017){Holland-Ashford}, {Lopez}, {Auchettl}, {Temim}, \& {Ramirez-Ruiz}}]{me17}
{Holland-Ashford}, T., {Lopez}, L.~A., {Auchettl}, K., {Temim}, T., \& {Ramirez-Ruiz}, E. 2017, \apj, 844, 84

\bibitem[{{Holland-Ashford} {et~al.}(2024){Holland-Ashford}, {Slane}, \& {Long}}]{me24}
{Holland-Ashford}, T., {Slane}, P., \& {Long}, X. 2024, \apj, 962, 82

\bibitem[{{Igoshev}(2020)}]{igoshev20}
{Igoshev}, A.~P. 2020, \mnras, 494, 3663

\bibitem[{{Janka}(2017)}]{janka17}
{Janka}, H.-T. 2017, \apj, 837, 84

\bibitem[{{Kaspi}(2000)}]{kaspi00}
{Kaspi}, V.~M. 2000, in Astronomical Society of the Pacific Conference Series, Vol. 202, IAU Colloq. 177: Pulsar Astronomy - 2000 and Beyond, ed. M.~{Kramer}, N.~{Wex}, \& R.~{Wielebinski}, 485

\bibitem[{{Katsuda} {et~al.}(2018){Katsuda}, {Morii}, {Janka}, {Wongwathanarat}, {Nakamura}, {Kotake}, {Mori}, {M{\"u}ller}, {Takiwaki}, {Tanaka}, {Tominaga}, \& {Tsunemi}}]{katsuda18}
{Katsuda}, S., {Morii}, M., {Janka}, H.-T., {et~al.} 2018, \apj, 856, 18

\bibitem[{{Kawata} {et~al.}(2019){Kawata}, {Bovy}, {Matsunaga}, \& {Baba}}]{gawa19}
{Kawata}, D., {Bovy}, J., {Matsunaga}, N., \& {Baba}, J. 2019, \mnras, 482, 40

\bibitem[{{Kim} {et~al.}(2024){Kim}, {Park}, {An}, {Mori}, {Reynolds}, {Safi-Harb}, \& {Zhang}}]{kim24}
{Kim}, C., {Park}, J., {An}, H., {et~al.} 2024, \apj, 977, 163

\bibitem[{{Lai}(2001)}]{lai01}
{Lai}, D. 2001, in Lecture Notes in Physics, Berlin Springer Verlag, Vol. 578, Physics of Neutron Star Interiors, ed. D.~{Blaschke}, N.~K. {Glendenning}, \& A.~{Sedrakian}, 424

\bibitem[{{Lee} {et~al.}(2020){Lee}, {Koo}, \& {Lee}}]{lee20}
{Lee}, Y.-H., {Koo}, B.-C., \& {Lee}, J.-J. 2020, \aj, 160, 263

\bibitem[{{Long} {et~al.}(2022){Long}, {Patnaude}, {Plucinsky}, \& {Gaetz}}]{long22}
{Long}, X., {Patnaude}, D.~J., {Plucinsky}, P.~P., \& {Gaetz}, T.~J. 2022, \apj, 932, 117

\bibitem[{{Lopez} {et~al.}(2009b){Lopez}, {Ramirez-Ruiz}, {Badenes}, {Huppenkothen}, {Jeltema}, \& {Pooley}}]{lopez09b}
{Lopez}, L.~A., {Ramirez-Ruiz}, E., {Badenes}, C., {et~al.} 2009b, \apjl, 706, L106

\bibitem[{{Lopez} {et~al.}(2009a){Lopez}, {Ramirez-Ruiz}, {Pooley}, \& {Jeltema}}]{lopez09a}
{Lopez}, L.~A., {Ramirez-Ruiz}, E., {Pooley}, D.~A., \& {Jeltema}, T.~E. 2009a, \apj, 691, 875

\bibitem[{{Mayer} {et~al.}(2020){Mayer}, {Becker}, {Patnaude}, {Winkler}, \& {Kraft}}]{mayer20}
{Mayer}, M., {Becker}, W., {Patnaude}, D., {Winkler}, P.~F., \& {Kraft}, R. 2020, \apj, 899, 138

\bibitem[{{Mayer} \& {Becker}(2021)}]{mayer21}
{Mayer}, M. G.~F., \& {Becker}, W. 2021, \aap, 651, A40

\bibitem[{{Narita} {et~al.}(2023){Narita}, {Uchida}, {Yoshida}, {Tanaka}, \& {Tsuru}}]{narita23}
{Narita}, T., {Uchida}, H., {Yoshida}, T., {Tanaka}, T., \& {Tsuru}, T.~G. 2023, \apj, 950, 137

\bibitem[{{Ranasinghe} {et~al.}(2019){Ranasinghe}, {Leahy}, \& {Tian}}]{ranasinghe19}
{Ranasinghe}, S., {Leahy}, D., \& {Tian}, W.~W. 2019, arXiv e-prints, arXiv:1910.05407

\bibitem[{{Reynolds} {et~al.}(2017){Reynolds}, {Pavlov}, {Kargaltsev}, {Klingler}, {Renaud}, \& {Mereghetti}}]{reynolds17}
{Reynolds}, S.~P., {Pavlov}, G.~G., {Kargaltsev}, O., {et~al.} 2017, \ssr, 207, 175

\bibitem[{{Scheck} {et~al.}(2006){Scheck}, {Kifonidis}, {Janka}, \& {M{\"u}ller}}]{scheck06}
{Scheck}, L., {Kifonidis}, K., {Janka}, H.-T., \& {M{\"u}ller}, E. 2006, \aap, 457, 963

\bibitem[{{Sett} {et~al.}(2021){Sett}, {Breton}, {Clark}, {van Kerkwijk}, \& {Kaplan}}]{sett21}
{Sett}, S., {Breton}, R.~P., {Clark}, C.~J., {van Kerkwijk}, M.~H., \& {Kaplan}, D.~L. 2021, \aap, 647, A183

\bibitem[{{Shan} {et~al.}(2018){Shan}, {Zhu}, {Tian}, {Zhang}, {Zhang}, {Wu}, \& {Yang}}]{shan18}
{Shan}, S.~S., {Zhu}, H., {Tian}, W.~W., {et~al.} 2018, \apjs, 238, 35

\bibitem[{{Shternin} {et~al.}(2019){Shternin}, {Kirichenko}, {Zyuzin}, {Yu}, {Danilenko}, {Voronkov}, \& {Shibanov}}]{shternin19}
{Shternin}, P., {Kirichenko}, A., {Zyuzin}, D., {et~al.} 2019, \apj, 877, 78

\bibitem[{{Temim} {et~al.}(2017){Temim}, {Slane}, {Plucinsky}, {Gelfand}, {Castro}, \& {Kolb}}]{temim17}
{Temim}, T., {Slane}, P., {Plucinsky}, P.~P., {et~al.} 2017, \apj, 851, 128

\bibitem[{Triggs {et~al.}(2000)Triggs, McLauchlan, Hartley, \& Fitzgibbon}]{triggs00}
Triggs, B., McLauchlan, P.~F., Hartley, R.~I., \& Fitzgibbon, A.~W. 2000, in Vision Algorithms: Theory and Practice, ed. B.~Triggs, A.~Zisserman, \& R.~Szeliski (Berlin, Heidelberg: Springer Berlin Heidelberg), 298--372

\bibitem[{{Tsuchioka} {et~al.}(2021){Tsuchioka}, {Uchiyama}, {Higurashi}, {Iwasaki}, {Otsuka}, {Yamada}, \& {Sato}}]{tsuchioka21}
{Tsuchioka}, T., {Uchiyama}, Y., {Higurashi}, R., {et~al.} 2021, \apj, 912, 131

\bibitem[{{T{\"u}llmann} {et~al.}(2010){T{\"u}llmann}, {Plucinsky}, {Gaetz}, {Slane}, {Hughes}, {Harrus}, \& {Pannuti}}]{tullmann10}
{T{\"u}llmann}, R., {Plucinsky}, P.~P., {Gaetz}, T.~J., {et~al.} 2010, \apj, 720, 848

\bibitem[{{Van Etten} {et~al.}(2012){Van Etten}, {Romani}, \& {Ng}}]{vanetten12}
{Van Etten}, A., {Romani}, R.~W., \& {Ng}, C.~Y. 2012, \apj, 755, 151

\bibitem[{{Verbunt} \& {Cator}(2017)}]{verbunt17}
{Verbunt}, F., \& {Cator}, E. 2017, Journal of Astrophysics and Astronomy, 38, 40

\bibitem[{{Vink}(2012)}]{vink12}
{Vink}, J. 2012, \aapr, 20, 49

\bibitem[{Virtanen {et~al.}(2020)Virtanen, Gommers, Oliphant, Haberland, Reddy, Cournapeau, Burovski, Peterson, Weckesser, Bright, {van der Walt}, Brett, Wilson, Millman, Mayorov, Nelson, Jones, Kern, Larson, Carey, Polat, Feng, Moore, {VanderPlas}, Laxalde, Perktold, Cimrman, Henriksen, Quintero, Harris, Archibald, Ribeiro, Pedregosa, {van Mulbregt}, \& {SciPy 1.0 Contributors}}]{scipy20}
Virtanen, P., Gommers, R., Oliphant, T.~E., {et~al.} 2020, Nature Methods, 17, 261

\bibitem[{{Weisskopf} \& {Hughes}(2006)}]{weiss06b}
{Weisskopf}, M.~C., \& {Hughes}, J.~P. 2006, {Six Years of Chandra Observations of Supernova Remnants}, ed. J.~W. {Mason}, 55

\bibitem[{{Winkler} {et~al.}(2009){Winkler}, {Twelker}, {Reith}, \& {Long}}]{winkler09}
{Winkler}, P.~F., {Twelker}, K., {Reith}, C.~N., \& {Long}, K.~S. 2009, \apj, 692, 1489

\bibitem[{{Wongwathanarat} {et~al.}(2013){Wongwathanarat}, {Janka}, \& {M{\"u}ller}}]{wongwathanarat13}
{Wongwathanarat}, A., {Janka}, H.-T., \& {M{\"u}ller}, E. 2013, \aap, 552, A126

\bibitem[{{Zeiger} {et~al.}(2008){Zeiger}, {Brisken}, {Chatterjee}, \& {Goss}}]{zeiger08}
{Zeiger}, B.~R., {Brisken}, W.~F., {Chatterjee}, S., \& {Goss}, W.~M. 2008, \apj, 674, 271

\bibitem[{{Zhou} {et~al.}(2020){Zhou}, {Zhou}, {Chen}, {Wang}, {Vink}, \& {Wang}}]{zhou20}
{Zhou}, P., {Zhou}, X., {Chen}, Y., {et~al.} 2020, \apj, 905, 99

\bibitem[{{Zhou} {et~al.}(2023){Zhou}, {Su}, {Yang}, {Chen}, {Sun}, {Jiang}, {Wang}, {Wang}, {Zhang}, {Xu}, {Yan}, {Yuan}, {Chen}, {Ao}, \& {Ma}}]{zhou23}
{Zhou}, X., {Su}, Y., {Yang}, J., {et~al.} 2023, \apjs, 268, 61

\end{thebibliography}

\end{document}